\documentstyle[12pt]{article}
\title{Suppressing Hawking radiation by quantum Zeno effect}
\author{Hrvoje Nikoli\'c \\
Theoretical Physics Division, Rudjer Bo\v{s}kovi\'{c} Institute, \\
P.O.B. 180, HR-10002 Zagreb, Croatia \\
{\normalsize e-mail: hnikolic@irb.hr} \\
\makebox[1in]{} \\
}
\date{\today}
\begin{document}
\maketitle
\begin{abstract}
We present evidence that quantum Zeno effect, otherwise working only for microscopic systems,
may also work for large black holes (BH's).
The expectation that a BH geometry should behave classically at time intervals larger than the Planck time
$t_{\rm Pl}$ indicates that the quantum process of measurement of classical degrees of freedom 
takes time of the order of $t_{\rm Pl}$. Since BH has only a few classical degrees of freedom,  
such a fast measurement makes a macroscopic BH strongly susceptible to the quantum Zeno effect,
which repeatedly collapses the quantum state to the initial one, the state before 
the creation of Hawking quanta. By this mechanism, Hawking radiation from a  BH of mass $M$ is strongly
suppressed by a factor of the order of $m_{\rm Pl}/M$.
\end{abstract}
\vspace*{0.5cm}
PACS Numbers: 04.70.Dy, 03.65.Ta \newline

\section{Introduction}

Fast-repeated measurements of an unstable quantum system 
may prevent its decay and thus stabilize it, by the mechanism known as
quantum Zeno effect \cite{misra,decoh1,zeno-review,auletta}.
As for most other quantum phenomena, it is typical for quantum Zeno effect that it
works for microscopic systems, not for the macroscopic ones.
 
In this paper, however, we present evidence that black hole (BH) 
may be an exception. BH geometry of a large black hole obeys classical laws,
while classicality is a consequence of fast measurement of classical properties, 
due to which a quantum superposition collapses
to a state in which classical observables have definite values.
Quantum mechanically, black hole is unstable owing to the Hawking radiation \cite{hawk1}.
We find that the quantum Zeno effect induced
by fast measurement of classical BH observables strongly suppresses creation of
Hawking radiation, with suppression being stronger when the black hole
is bigger.

Of course, staring at a macroscopic piece of material containing $10^{23}$ atoms in unstable excited 
states will not stop the decay of atoms. Essentially, this is because different atoms
decay independently, and in practice one cannot monitor $10^{23}$ independent degrees of freedom.
In the laboratory, quantum Zeno effect works if the observation is applied
to a single atom, or at best to a relatively small number of them. 

But how then the quantum Zeno effect may work for the macroscopic black hole,
having a large number of degrees of freedom? Indeed, the number of degrees of freedom 
may be counted by the BH entropy 
\begin{equation}\label{e1}
S\sim A/l_{\rm Pl}^2 \sim R^2/l_{\rm Pl}^2 \sim M^2/m_{\rm Pl}^2 , 
\end{equation}
where $A$ is the BH surface, $R$ is the BH radius, $M\sim R m_{\rm Pl}^2$ is the BH mass, 
$l_{\rm Pl}$ is the Planck length, $m_{\rm Pl}=1/l_{\rm Pl}$ is the Planck mass, and
we use units in which Planck constant, velocity of light, and Boltzmann constant 
are set to unity: $\hbar=c=k_{\rm B}=1$. The entropy (\ref{e1}) is very large for macroscopic
black holes with $M\gg m_{\rm Pl}$, so at first sight it may seem that quantum Zeno effect
cannot work for macroscopic black holes.

However, most of these degrees of freedom are irrelevant for Hawking radiation.
Instead, Hawking radiation depends only on a few classical degrees of freedom \cite{hawk1};
mass, charge, and angular momentum of the black hole {\em as a whole}. In this sense Hawking
radiation is {\em not} like a radiation from a box containing many independent atoms in excited 
states. Instead, it is more like a radiation from a {\em single} atom. 
But at the same time, a typical black hole is much bigger than an ordinary atom, so it may interact 
with a much larger number of environment degrees of freedom. The interaction with a larger number
of environment degrees implies a faster process of decoherence 
and hence a faster process of measurement \cite{decoh1,decoh2}, which strongly suggests that
the measurement of the relevant BH degrees of freedom is much faster than for ordinary atoms.
In this way, the remarkable BH property of being both 
macroscopic (by size and mass) and microscopic (by the small number of classical degrees of freedom)  
makes black holes extremely susceptible to the quantum Zeno effect.

In the rest of the paper we put these qualitative arguments into a more quantitative form.    
In the absence of a complete quantum theory of gravity, however, a lot of
quantitative arguments will rely on the order-of-magnitude estimates. 

\section{Quantum Zeno effect}

Let us first outline how the quantum Zeno effect works
for ordinary quantum systems \cite{misra,decoh1,zeno-review,auletta}. 
Initially, let the system be in an unstable state $|\Psi_0\rangle$, and 
let $P_0(t)$ denote the survival probability of remaining in the initial state.
For a sufficiently long time $t$, 
the survival probability $P_0(t)$ obeys the exponential law
$P_0(t)=P_{\gamma}(t)$, where
\begin{equation}\label{e2}
P_{\gamma}(t) \equiv e^{-\gamma t} .
\end{equation}

This exponential law is valid when there are no measurements before $t$, but how can measurements change it? 
For simplicity, suppose
that during the time $t$ there was $N$ instantaneous measurements occurring at times $nt/N$,
$n=1,\ldots,N$, with equal periods $t/N$ of free evolution between the measurements.
If the time $t/N$ is sufficiently long so that the exponential law (\ref{e2}) is still
valid, then the survival probability after time $t$ is 
$P_0(t)=[P_{\gamma}(t/N)]^N=e^{-\gamma t}$, which does not differ from (\ref{e2}).

But what if $t/N$ is so small that the exponential law is not valid? The expansion of (\ref{e2})
for small $t$ leads to a {\em linear} law
\begin{equation}\label{e3}
P_{\gamma}(t)\simeq 1-\gamma t ,
\end{equation}
so with $N$ measurements the survival probability is
\begin{equation}\label{e4}
P_0(t)=(1-\gamma t/N)^N \; \stackrel{N\rightarrow\infty}{\longrightarrow} \; e^{-\gamma t} ,
\end{equation}
naively suggesting that the exponential law is valid even for very large $N$, corresponding 
to the very small $t/N$.

However, the linear law (\ref{e3}) for very short times is actually wrong. (The short-time expansion
of a function valid for long times does not need to lead to a correct result for short times.)
To find the correct law, one must start from first principles. If $H$ is the relevant Hamiltonian, 
including the interaction which makes the system in the initial state $|\Psi_0\rangle$ unstable, 
then the state at time $t$ has the form
\begin{equation}\label{e5}
|\Psi(t)\rangle=e^{-iHt}|\Psi_0\rangle =c_0(t)|\Psi_0\rangle + 
\sum_{k\neq 0} c_k(t)|\Psi_k\rangle , 
\end{equation}
where
$c_k(t)=\langle \Psi_k|\Psi(t)\rangle =\langle \Psi_k|e^{-iHt}|\Psi_0\rangle$,
and each $|\Psi_k\rangle$ for $k\neq 0$ is a possible decayed state. Thus the short-time
expansion gives
\begin{eqnarray}\label{e7}
c_0(t) &=& \langle \Psi_0|e^{-iHt}|\Psi_0\rangle 
\nonumber \\
&\simeq& 1-it\langle \Psi_0|H|\Psi_0\rangle -\frac{t^2}{2}\langle \Psi_0|H^2|\Psi_0\rangle ,
\end{eqnarray}
so
\begin{equation}\label{e8}
P_0(t)=c_0^*(t)c_0(t)\simeq 1-t^2 (\Delta H)^2 ,  
\end{equation} 
where
\begin{equation}\label{e9H}
(\Delta H)^2=\langle \Psi_0|H^2|\Psi_0\rangle - \langle\Psi_0|H|\Psi_0\rangle^2 .
\end{equation}
The uncertainty of energy $\Delta H$ is non-zero whenever $|\Psi_0\rangle$ is unstable,
for if the uncertainty were zero, then $|\Psi_0\rangle$ would be an eigenstate of the  
Hamiltonian $H$ and hence would be stable. 

Unlike (\ref{e3}), (\ref{e8}) is {\em quadratic} in time! Hence, if now we introduce
measurements, the survival probability is not given by (\ref{e4}), but by   
\begin{equation}\label{e10}
P_0(t)=[1- (\Delta H)^2 (t/N)^2]^N \simeq 1-N(\Delta H)^2 (t/N)^2 .
\end{equation}
We see that $\lim_{N\rightarrow\infty} P_0(t) =1$, so in this limit the initial state 
$|\Psi_0\rangle$ survives
with {\em certainty}. In other words, a continuous measurement prevents the decay
of the system, which otherwise would decay if the measurement was absent.
This is the quantum Zeno effect. 

Of course, a realistic measurement cannot be perfectly continuous because each
measurement lasts a finite time $t_{\rm meas}$, given by the time
needed for the process of decoherence \cite{decoh1,decoh2}. So in practice,
the quantum Zeno effect is efficient when $t_{\rm meas}$
is much shorter than the Zeno time 
\begin{equation}\label{e11}
t_{\rm Zeno}=1/\Delta H.
\end{equation}

Let us recapitulate the main assumptions that were used to get an efficient quantum Zeno effect.
First, the number of degrees of freedom relevant for the quantum decay must be small.
(Otherwise, it is hard to measure them.)
Second, the short-time evolution of the system in the absence 
of measurement is a unitary evolution governed by a time-independent Hamiltonian.
Third, the measurement must be sufficiently fast, so that
\begin{equation}\label{e12Z}
t_{\rm meas} \ll t_{\rm Zeno}.
\end{equation}  
In the following we shall explain how all three assumptions get satisfied for 
black holes.

\section{Number of relevant degrees of freedom}

Owing to the classical BH no-hair theorems 
\cite{hawking-ellis,frolov-novikov},
an arbitrary initial black hole soon settles down to a stationary black hole
characterized by only a few degrees of freedom: mass, charge, and angular momentum.
(All other classical degrees of freedom are radiated away by gravitational waves.) 
Hawking radiation, derived from the assumption of a stationary classical BH background \cite{hawk1},
depends only on these three quantities. Thus, these three classical degrees of freedom
are the only BH degrees of freedom which are relevant for Hawking radiation. 

The fact that the black hole contains only a few relevant degrees of freedom can also be confirmed
by a semi-classical thermodynamic argument. For that purpose, let us temporarily ignore 
the quantum Zeno effect. Then, far from the horizon, the Hawking radiation makes the black hole look like 
a black body of volume $V=4\pi R^3/3$ having the Hawking temperature \cite{hawk1}
\begin{equation}\label{e12}
T_{\rm H}=\frac{m_{\rm Pl}^2}{8\pi M} .
\end{equation}
To give an effective description of this radiation, one may ignore gravity and model black hole as a box of 
volume $V$ filled with nothing but free Hawking particles at temperature $T=T_{\rm H}$. 
The corresponding effective BH entropy is given by standard non-gravitational statistical physics \cite{kapusta}
\begin{equation}\label{e13}
S_{\rm eff}=\frac{\partial(T \ln Z)}{\partial T} ,
\end{equation}      
where (for a single particle species)
\begin{equation}\label{e14}
\ln Z = Vg\int\frac{d^3p}{(2\pi)^3}\frac{1}{1- e^{-\omega/T}} , 
\end{equation}
$\omega=\sqrt{{\bf p}^2+m^2}$, and $g\sim 1$ is a spin factor. 
Hawking radiation consists mainly of massless particles, so we can take $m=0$. 
Therefore the integral in (\ref{e14}) is of the order of $T^3$, so (\ref{e13}) with $T=T_{\rm H}$ gives
\begin{equation}\label{e15S}
S_{\rm eff}\sim VT^3 \sim R^3 T_{\rm H}^3 \sim (M/m_{\rm Pl}^2)^3 (m_{\rm Pl}^2/M)^3 =1 .
\end{equation}
The fact that this effective entropy is of the order of unity
confirms that only a few degrees of freedom are relevant for Hawking radiation.

Note that (\ref{e15S}) is much smaller than (\ref{e1}), which can be easily understood
from the BH thermodynamic relation $dS=dM/T$ and the fact that the BH mass $M$
can {\em not} be modeled by a box containing nothing but free Hawking particles at temperature $T_{\rm H}$.
Nevertheless, such an effective model is appropriate for a description of thermal black-body 
{\em radiation}. As known from standard non-gravitational statistical physics, 
two black bodies with the same $T$ and $V$ radiate equally,
even if they have very different masses or thermal energies. 

\section{Unitarity and measurement}

Hawking radiation is in a possible conflict with unitarity
for very long times, during which the black hole suffers a significant decrease of mass
\cite{hawk2,gid,math,hoss,fabbri}. 
But for shorter times,
the description based on Bogoliubov transformation is fully unitary \cite{sidorov}, owing to the
entanglement between outside an inside Hawking particles. In addition, all treatments of Hawking radiation
take for granted that backreaction provides energy conservation. This short-time unitarity and energy conservation 
imply that the short-time evolution can be described by (\ref{e5}) with an
effective time-independent Hamiltonian $H=H_{\rm eff}$.
Similarly to the effective entropy (\ref{e15S}),  
the effective Hamiltonian does not describe all degrees of freedom, but only those which are relevant for
Hawking radiation.
The states $|\Psi_k\rangle$ in (\ref{e5}) can be chosen to be the free-Hamiltonian
eigenstates \cite{nik-epjc} 
\begin{equation}\label{e15}
|\Psi_k\rangle = |M-E_k\!> |E_k\rangle ,
\end{equation}
where $|E_k\rangle$ is a state of Hawking radiation with energy $E_k$,
$M$ is the initial BH mass, and $|M-E_k\!>$ is a BH state with mass $M-E_k$. 
Likewise, $|\Psi_0\rangle = |M\!> |0\rangle$. For relatively long times (but still
short enough so that the BH temperature has not changed much), 
$c_k\propto e^{-E_k/2T_{\rm H}}$ \cite{nik-epjc}.    

The state (\ref{e5}) with (\ref{e15}) involves a superposition of BH states with different values of masses.
But a classical BH state has a definite mass. Since macroscopic BH geometry behaves 
classically, this means that (\ref{e5}) ``collapses" to one of the states with a definite mass.
The ``collapse" is caused by measurement, which can be viewed as entanglement with the environment degrees 
of freedom which perform the measurement \cite{decoh1,decoh2}. This means that (\ref{e15}) modifies to
\begin{equation}
|\Psi_k\rangle = |M-E_k\!> |{\cal E}_{M-E_k}\!> |E_k\rangle , 
\end{equation}
where $|{\cal E}_{M'}\!>$ is a 
state of environment corresponding to a situation in which the BH mass is measured to have the value $M'$.
In particular, $|{\cal E}_{M'}\!>$ contains information about classical geometry surrounding the black hole.
A more complete description of measurement involves also a measurement of BH charge $Q$ and angular
momentum $J$, leading to the environment states of the form $|{\cal E}_{M'Q'J'}\!>$.     

\section{The small-energy resolution problem}
\label{SECresolution}

The presence of environment degrees of freedom which measure the black hole
must be compatible with the fact that the
classical Schwarzschild black hole is time-independent in standard Schwarzschild coordinates.
This will be fulfilled if the environment degrees of freedom are static themselves, i.e. do not move
with respect to the black hole. Furthermore, the locality of interactions between the black hole
and its environment implies that all measurements can be thought of as being performed at well
defined distances from the horizon.
Hence, all measurements by the environment are assumed to correspond to an ``observer'' which is
static with respect to the black hole and sits at a well defined distance from the horizon.  

For the quantum Zeno effect to work, 
the measurement of BH mass must be able to distinguish 
different BH masses. This means that the environment state $|{\cal E}_{M-E_k}\!>$ must be sufficiently
different from the state $|{\cal E}_{M}\!>$. But typical $E_k$ is of the order of $T_{\rm H}$, 
which is a very small energy. Far from the horizon, such small differences of the BH mass 
cannot be resolved. 

Nevertheless, they can be resolved near the horizon. This is because
a Hawking particle, having a small red-shifted energy $E_k\sim T_{\rm H}$ far from the horizon, has 
a large trans-planckian energy close to the horizon. 
We assume that the particle is created near horizon as a wave packet \cite{hawk1} 
with uncertainty of energy and momentum
comparable to its average energy and momentum, implying that the particle near horizon
with a large average energy $E$ is located within a small length of the order of $1/E$. 
Hence, assuming that measurement 
makes geometry classical even close to the horizon, different states in (\ref{e5})
can be resolved despite the fact that the temperature measured far from the horizon is very small. 

At first sight, one might think that the assumption of classical geometry near horizon
contradicts expectations from various approaches to quantum gravity. Nevertheless there
is no contradiction, as long as the ``near horizon'' region corresponds to a distance
which is still much larger than the Planck length $l_{\rm Pl}$. Indeed, the metric of
the Schwarzschild black hole can be written as
\begin{equation}
 g_{00}=\frac{-1}{g_{rr}}=1-\frac{R}{r} ,
\end{equation}
where the Schwarzschild radius
\begin{equation}
 R=2l_{\rm Pl} \frac{M}{m_{\rm Pl}}
\end{equation}
satisfies $R \gg l_{\rm Pl}$ for the macroscopic black hole with $M \gg m_{\rm Pl}$.
The ``near horizon'' region corresponds to the distance $r$ which satisfies
\begin{equation}
 \frac{r-R}{R} \ll 1 ,
\end{equation}
so the small distance from the horizon $r-R\equiv \Delta r$ must satisfy
\begin{equation}\label{rR}
 \Delta r \ll R . 
\end{equation}
Hence, despite of being ``small'', the distance $\Delta r$
may in fact be much larger than the Planck length $l_{\rm Pl}$.
Even though many approaches to quantum black holes (e.g. loop quantum gravity \cite{rov},
gauge/gravity duality \cite{gauge-gravity}, fuzzballs \cite{math}, or firewalls \cite{AMPS})
predict that a black hole seen by a static 
observer siting at a Planck distance from the horizon is highly non-classical,
they typically do not predict significant violation of classicality at distances compatible with (\ref{rR})
which are much larger than the Planck length.  

We have argued that radiation from a black hole is similar to radiation from a single atom.
However, there is one important difference between atoms and black holes,
that works against efficiency of the quantum Zeno effect
for black holes. Unlike radiation from an atom, Hawking radiation from a macroscopic black hole
has a continuous energy spectrum. In particular, the energy of the Hawking particle can be arbitrarily small.
Even close to the horizon, the Hawking particle may be created with an arbitrarily small energy.
Such arbitrarily small energies cannot be resolved by the measurement, implying that
quantum Zeno effect cannot stop creation of Hawking particles with such small energies
near the horizon. This certainly decreases the efficiency of the quantum Zeno effect. 

Nevertheless, this does not change the fact that the quantum Zeno effect
suppresses Hawking radiation significantly. Namely, Hawking particle created with such a small 
energy near the horizon will have an even smaller energy far from it. In this way, particles
for which the quantum Zeno effect will not work will escape with energies $E_k\ll T_{\rm H}$. 
As a consequence, even though there will be some radiation from the black hole,
the intensity of radiation will be much smaller than predicted by the standard Hawking analysis.

\section{Typical time scales}

What is the typical time scale $t_{\rm meas}$ needed to perform a 
measurement of classical BH observables such as mass? To answer that question, in principle
one would need a complete theory of quantum gravity. But even in the absence of such a theory, 
it is not difficult to estimate the order of magnitude. It is widely expected 
that quantum theory of gravity can be approximated by the classical theory for times
longer than the Planck time $t_{\rm Pl}=1/m_{\rm Pl}$. On the other hand, we have seen that classicality 
of gravity is a consequence of measurement of the classical gravitational observables. 
Hence
\begin{equation}\label{e16}
t_{\rm meas} \sim t_{\rm Pl} .
\end{equation}  

%
We also stress that the black hole is {\em not} classical at time scales shorter or equal to
(\ref{e16}). During the short time (\ref{e16}) the black hole can be thought of as continuously evolving
from a quantum superposition of different masses to a classical state with a definite mass,
becoming a state with a definite mass only after a time longer than (\ref{e16}).
This seems to be qualitatively compatible with other approaches to quantum black holes 
mentioned in Sec.~\ref{SECresolution}, which typically predict significant deviations
from classical gravity at short space and time scales of the order of Planck scale,
but not at longer scales.

The very short time (\ref{e16}) is what is needed for the quantum Zeno effect to work.
More precisely, the quantum Zeno effect needs the condition (\ref{e12Z}). Thus we need to show
that $t_{\rm Pl}$ is much shorter than $1/\Delta H$ in (\ref{e11}), so
we need to determine the value of $\Delta H$. Since $H=H_{\rm eff}$ is the effective Hamiltonian describing
only those degrees of freedom which are relevant for Hawking radiation, $\Delta H$
can be estimated by an analysis similar to the one which we used to obtain
(\ref{e13})-(\ref{e15S}). By viewing black hole as a box filled with free particles at temperature $T$,
the uncertainty $\Delta H$ can be identified with the uncertainty of energy due to the
thermal fluctuations inside the volume $V$. The thermal average of the $n$'th power of thermal energy is \cite{kapusta}
\begin{equation}\label{e17}
\langle E^n\rangle = Vg\int\frac{d^3p}{(2\pi)^3}\frac{\omega^n}{e^{\omega/T} - 1} .
\end{equation}
The integral in (\ref{e17}) for massless particles is of the order of $T^{3+n}$, so
\begin{eqnarray}\label{e18}
\langle E^n\rangle &\sim & V T^{3+n} \sim R^3 T_{\rm H}^{3+n} 
\nonumber \\
&\sim & (M/m_{\rm Pl}^2)^3 (m_{\rm Pl}^2/M)^{3+n} = (m_{\rm Pl}^2/M)^n . 
\end{eqnarray}    
Hence
$\langle E^2\rangle \sim \langle E\rangle^2 \sim (m_{\rm Pl}^2/M)^2$, 
so
\begin{equation}\label{e19}
(\Delta E)^2 = \langle E^2\rangle - \langle E\rangle^2 \sim (m_{\rm Pl}^2/M)^2 .
\end{equation}
Therefore (\ref{e11}) is estimated to be
\begin{equation}\label{e20}
t_{\rm Zeno} \sim \frac{1}{\Delta E} \sim \frac{1}{m_{\rm Pl}} \frac{M}{m_{\rm Pl}} = 
t_{\rm Pl}\frac{M}{m_{\rm Pl}}  ,
\end{equation}
which is much larger than (\ref{e16}). This shows that the condition (\ref{e12Z}) is fulfilled.

For a comparison, we also need to know the time scale 
\begin{equation}\label{t-gamma}
t_{\gamma}=1/\gamma
\end{equation}  
in (\ref{e2}). Since this time is relevant when the quantum Zeno effect is not present,
we determine it by considering Hawking radiation in the absence of quantum Zeno effect. 
For long times, Hawking radiation can be described as a continuous process,
during which the black hole looses mass at the rate \cite{fabbri,mukhanov}  
\begin{equation}\label{dM}
\frac{dM}{dt}\sim -\frac{m_{\rm Pl}^4}{M^2} . 
\end{equation}
But at shorter times the radiation is better viewed as a series of
discrete quantum jumps, where black hole looses mass 
of the order $\Delta M\sim -T_{\rm H}$ in a typical jump. Hence we write (\ref{dM}) in a discretized form
\begin{equation}\label{DM}
\Delta t \sim -\Delta M \frac{M^{2}}{m_{\rm Pl}^{4}} \sim T_{\rm H} \frac{M^{2}}{m_{\rm Pl}^{4}} ,
\end{equation} 
which is the typical time needed for one jump to occur. When a jump occurs then
the system ceases to be in the initial state, so from (\ref{e2}) and (\ref{t-gamma}) we see 
that $t_{\gamma}\sim \Delta t$. Therefore 
\begin{equation}\label{t-gamma2}
t_{\gamma} \sim T_{\rm H} \frac{M^{2}}{m_{\rm Pl}^{4}}  
\sim \frac{m_{\rm Pl}^2}{M} \frac{M^{2}}{m_{\rm Pl}^{4}} = t_{\rm Pl}\frac{M}{m_{\rm Pl}} ,
\end{equation} 
which is the same order of magnitude as (\ref{e20}).

\section{Suppression of Hawking radiation}

Now we can finally estimate how much the Hawking radiation
is suppressed by the quantum Zeno effect. The probability that decay will happen during the time $t$ is
\begin{equation}\label{P(t)}
p(t)=1-P_0(t).
\end{equation}
In the absence of quantum Zeno effect this is determined by the value of $\gamma$ in (\ref{e2}),
so (\ref{P(t)}) for short times is equal to
\begin{equation}\label{Pgamma}
p_{\gamma}(t)\simeq \gamma t =\frac{t}{t_{\gamma}} \sim  \frac{m_{\rm Pl}}{M} \frac{t}{t_{\rm Pl}}.
\end{equation}
On the other hand, when the quantum Zeno effect is present then $P_0(t)$ is given by (\ref{e10}), 
so in this case (\ref{P(t)}) is equal to
\begin{equation}\label{Pzeno}
p_{\rm Zeno}(t)\simeq (\Delta H)^2 t^2 /N.
\end{equation}
The number of measurements at time $t$ is
\begin{equation}
N=\frac{t}{t_{\rm meas}} \sim \frac{t}{t_{\rm Pl}} ,
\end{equation}
while $\Delta H\sim \Delta E$ is given by (\ref{e19}). Therefore (\ref{Pzeno}) is
\begin{equation}\label{Pzeno2}
p_{\rm Zeno}(t)\sim \left( \frac{m_{\rm Pl}}{M} \right)^2 \frac{t}{t_{\rm Pl}} .
\end{equation}

The strength of the suppression by the quantum Zeno effect is given by the ratio between 
the decay probability (\ref{Pzeno2}) with the Zeno effect and the decay probability (\ref{Pgamma}) 
without the Zeno effect.
The results above show that this ratio is 
\begin{equation}\label{Gamma}
\Gamma \equiv \frac{p_{\rm Zeno}(t)}{p_{\gamma}(t)} \sim \frac{m_{\rm Pl}}{M} .
\end{equation}
This is a very small number for a macroscopic black hole with $M\gg m_{\rm Pl}$, showing that 
the suppression of radiation by the quantum Zeno effect is very strong. Moreover, 
the suppression is stronger when $M$ is larger, i.e., when the black hole is bigger. 
In contrast with the usual expectations about quantum phenomena, 
{\em the more macroscopic black hole is, the more efficient quantum Zeno effect becomes.}

We have also seen that quantum Zeno effect is not efficient for particles which
escape with sufficiently small energy $E_k\ll T_{\rm H}$. This means that $\Gamma$ is not a constant,
but a function $\Gamma(E_k)$ with the properties $\Gamma(0)=1$, 
$\Gamma(E_k)\sim m_{\rm Pl}/M$ for  
$E_k\:\, {\raise-.5ex\hbox{$\buildrel{{\textstyle \!>}}\over{\!\sim}$}} \,T_{\rm H}$.

The result $\Gamma(E_k)\neq 1$ means that radiation from the black hole is not 
a black-body radiation at temperature $T_{\rm H}$. Instead, $\Gamma(E_k)$ can be viewed
as a grey-body factor, which modifies BH radiation in a way similar to the grey-body 
factors coming from other physical mechanisms (see e.g. \cite{hawk1,fabbri,mukhanov}).    

\section{Conclusion}

Typically, quantum Zeno effect is a microscopic effect,
working when a small number of relevant degrees of freedom of an unstable system is 
frequently measured, due to which the decay of the unstable system is suppressed.
We have presented evidence that black hole, being a macroscopic system with only a few
degrees of freedom relevant for Hawking radiation, is a macroscopic system 
for which the quantum Zeno effect is very strong. The measurement of classical
observables of a macroscopic black hole at the time scale of the order of Planck time
drives the quantum Zeno effect, which strongly suppresses creation of Hawking radiation. 
    
\section*{Acknowledgements}

The author is grateful to an anonymous referee for useful suggestions 
to improve the paper. 
This work was supported by the Ministry of Science of the
Republic of Croatia under Contract No.~098-0982930-2864.

\end{document}